# The Calculus of Democratization and Development

Jacob R. Ferguson

Michigan State University

Author Note

Numerous graphs for all countries in the large N set were produced, and can be found in the accompanying Excel sheet.

Abstract

In accordance with "Democracy's Effect on Development: More Questions than Answers", we seek to carry out a study in following the description in the 'Questions for Further Study.' To that end, we studied 33 countries in the Sub-Saharan Africa region, who all went through an election which should signal a "step-up" for their democracy, one in which previously homogenous regimes transfer power to an opposition party that fairly won the election. After doing so, liberal-democracy indicators and democracy indicators were evaluated in the five years prior to and after the election took place, and over that ten-year period we examine the data for trends. If we see positive or negative trends over this time horizon, we be able to conclude that it was the recent increase in the quality of their democracy which lead to it. Having investigated examples of this in depth, there seems to be three main archetypes which drive the results. Countries with positive results to their democracy from the election have generally positive effects on their development, countries with more "plateau" like results also did well, but countries for whom the descent to authoritarianism was continued by this election found more negative results.

*Keywords*: Democracy, Development, V-Dem, WDI, World Bank, Calculus, Mixed-N

The Calculus of Democratization and Development

In investigating the causal relationship between democracy and development, a field of mathematics which seems likely to contain answers the answers we need would be calculus. In calculus, we are primarily concerned with the *change in* a given variable over another variable. For us those variables are our democracy and development indicators, and the other variable will always be time. To that end, when we examine the derivative of the linear approximation of the data, at the point we care about, the election, should be able to use that information to make some sort of assertion about the causal link between democracy and development, or the lack thereof. In our investigation we have found that democracies undergo this democratic transition in a number of archetypal modes. The three most important found in our investigation were “Election Plateau,” “Slope Down,” and “Step Up.” Amongst these we selected one country from each of those three archetypes to examine in detail, Niger, Benin, and Zimbabwe.

## Methodology

In the analysis of the countries of the SSA region, we first collected the relevant variables for examination from the World Bank’s WDI dataset and V-Dem’s datasets. Then, once all the data we would need was properly added, we continued by identifying countries which underwent their first peaceful democratic transition of executive power after, on average, 1963. Next, the countries had their democracy and development indices evaluated over time, specifically the 5 years before and 5 years after the key election. Then, a linear regression was applied to the data, and that linear regression was used to determine the trend about the transition date. Originally, we expected to find that all the countries in our study would exhibit the “step up” archetype, but as it turns out, sometimes, those transitions either empowered new authoritarians or did not affect long term change, as the newly elected government was later removed in some sort of authoritarian coup.

**Literature Review**

In examining the current state of consensus on this issue amongst political scientists, it is imperative that we better restrict our focus specifically to the casual links existence and directionality, as this is the area we are most concerned with in this study. According to Olson's 1993 piece, "Dictatorship, Democracy, and Development," while authoritarianism is preferable to anarchy, the individual rights guaranteed by a strong, sustaining, liberal democracy are necessary to maximize private investment and economic growth. Clearly then, it seems that Olsen is amenable to the idea that democracy is a causal factor in development in countries, especially when it comes to seeing countries reach their full economic and quality of life potential. Shihata's 1997 piece, "Democracy and Development" again sheds light on this subject, summarizing much of the World Bank's observations on the matter, which seem to find it difficult to qualify claims of causality in either direction. Hopefully, this is where my research can shed some light on the subject. Still, even Shihata must eventually concede that, more often than not, democracies show stronger development. The main outlier amongst the literature reviewed was certainly the piece by Przeworski, Limongi, and Giner, "Political Regimes and Economic Growth." In this piece, they attack the issue from a different position, they begin at a simple point of agreement amongst the majority of the political science community; property rights are critical and causal to development and growth in a country. Then, they proceed via a new question, "Does democracy or authoritarianism better provide for property rights?" Surprisingly, and unfortunately for a pro-democracy point of view, Przeworski et al. found that Authoritarianism was better at properly insulating against the pressure for immediate consumption, which would hinder overall investment. While one can easily envision possible ways to refute the findings of Przeworski et al, even if their findings are true as long as the other benefits of democracy outweigh the negative effects on property rights that Przeworski et al. cite, then we would still expect our findings to consistent with the majority, that democracy is generally beneficial to development. While most of the papers reviewed hinted at the democracy causing development link, nearly all acknowledged that the other direction was certainly also existent. Often, authors made mention of the need to reach certain development thresholds to be ready for a democracy, especially a sustainable democracy (Shihata 1997). Overall, while it seems clear that the fact that development is often necessary for democracy, it is the question of whether democracy leads to development that is currently the most ripe and ready for examination, since consensus on the topic is mostly weak.

## Large N – the 33 SSA Countries

As we discussed above all previously examined SSA countries were evaluated for whether they would have a first ever election turnover in a time frame in which sufficient WDI data existed to conduct our analysis. That left us with a large N of 33 suitable countries, some of which became insignificant later in the investigation. Amongst those we began to divide the countries up by the response their liberal democracy index had to the election. While numerous archetypes existed, “Election Plateau,” “Election Spike,” “Hockey Stick,” “Slope Down,” “Slope Up,” “Step Down,” Step Up,” only three were selected as important enough for further study. Those three were “Election Plateau,” “Slope Down,” and “Step Up,” and each represents a different story, and result for authoritarianism and democracy in their country.

**Election Plateau**

The Election Plateau is the most interesting of the three archetypes we found in our study. While one might expect that these power transferring elections would result in increases in the liberal democracy indexes of those countries involved, hat we found instead was that for some countries, the election was just the exchange of one parties chosen dictator for another. For these countries, while they often experienced overall GDP losses, often these new regimes were able to rally the country to new GDP growth heights and better international trade. These results give our conclusions the greatest trouble, because while the election represented a surge in liberal democracy, it appears that after that surge the country retreated back to authoritarianism, without most of the detriment that might seem to imply. It may likely take extensive further investigation whether the positive effects were felt because of the democratic surge, or from the stabilizing retreat of liberal democracy.

**Slope Down**

The slope down indicates a dismissal of liberal democracy before, during, and after the key election. The development of country seems untenable by the administration that continues, unchanged, the countries path to authoritarianism. One of the most interesting trends identified, was that when the GDPs of slope down countries showed a negative trend, their GDP growth was positive. And, when the GDP of those countries trends positive, their GDP growth suffered. This would seem to imply that when an authoritarian regime takes control of a strong economy turning bad, it is within their power to improve it. But, when inheriting a weak economy experiencing a growth, they seem incapable of making that better. This is promising if, as democrats, we would like to see authoritarian regimes be less competent in effecting positive change in development factors than democracies.

**Step Up**

The step up countries exemplify what we should hope to see in the event of a true democratic transition. The election represented a clear slope up in the liberal democracy index of the country, where the country then stabilizes at a higher level of liberal democracy. Luckily for us, our results seem conducive to the idea that democracies promote development. Step up countries tend to show positive trends in GDP and GDP growth. More often than not, they also see positive trends in primary education completion amongst relevant age groups. This one of the hall mark findings of our study, because while counter examples exist, the majority of step up countries show marked improvements in development as a result of this step up in liberal democracy. This does open up further questions, but give us the best most direct answer to the question between democracy and development to date.

## Small N Analysis

In following up our examination of the three archetypes we select three key cases to further examine to better understand the phenomena underlying the democratic transitions in those nations. To that end, we select the countries of Niger, Benin, and Zimbabwe. Niger exemplifies the election plateau archetype, showing off how despite a negative GDP trend, all other variables analyzed experienced positive reactions to the temporary boost in liberal democracy indices. Benin exemplifies the slope down archetype, where having inherited a weak economy growing stronger, the GDP growth trend is negative, implying that the administration is failing to secure actual growth to policy. Zimbabwe, exemplifies the step up archetype, in which strong growth of in real GDP per capita and in the growth rate of their GDP shows honest improvements. Zimbabwe's development improvements speak to their improved liberal democracy and how it leads to greater development.

**Niger**

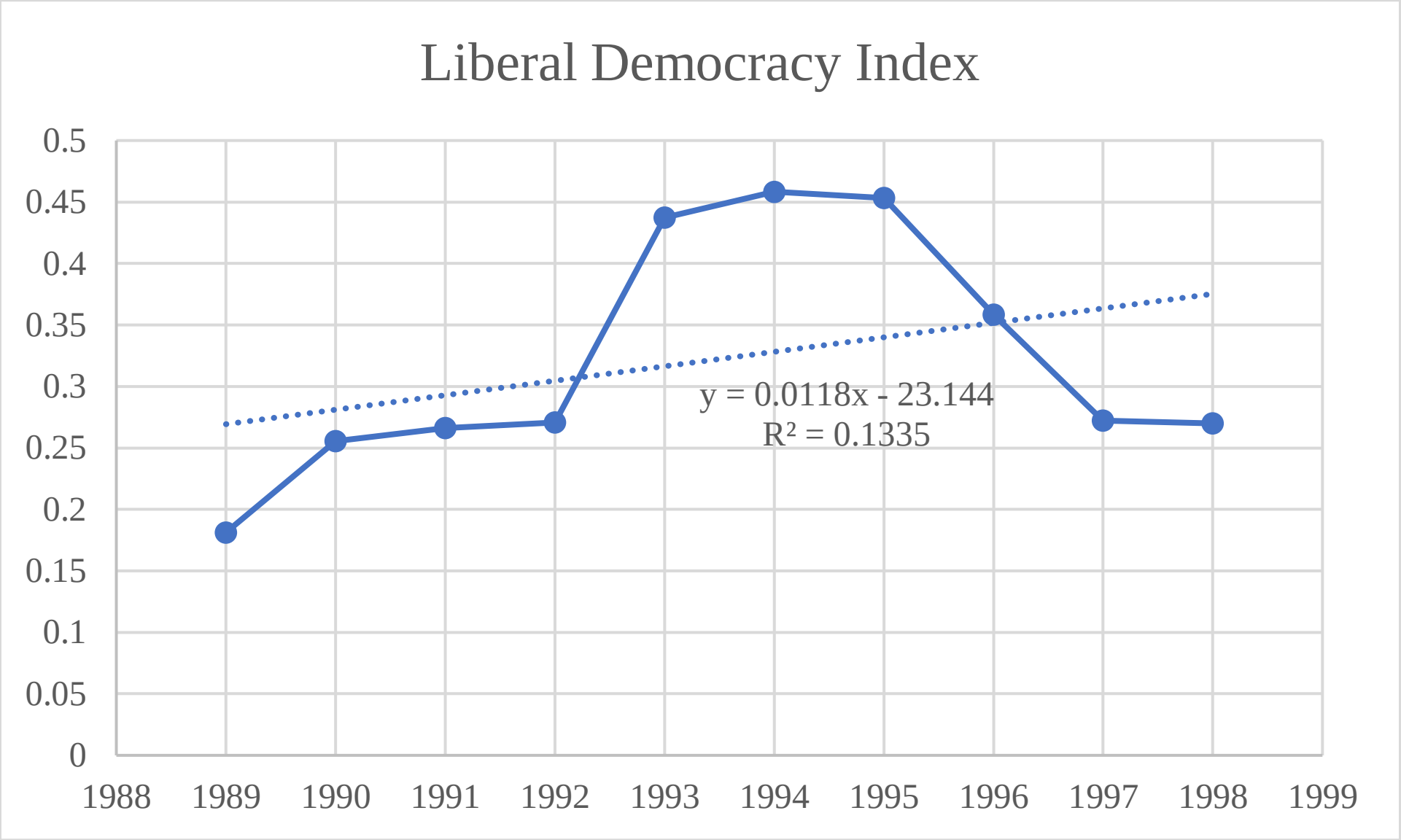

*Figure 1.1*

Niger shows how under an election plateau condition a country can experience an overall growth in GDP per capita, but suffer from a negative trend in GDP growth. As we have previously seen, while more democratic countries tend to have more development, the extenuating circumstances of election plateau countries cast that knowledge into doubt. Niger had troubles from the start, from its independence in 1958, it struggled to gain a solid democratic standing. In its first 14 years, it existed as a single party civilian regime under the then president, Hamani Diori (Idrissa & Decalo 2012). Then, the military, displeased with the actions of the ruling regime in response to an ongoing drought along with plentiful accusations of corruption, seized power through a coup and instated the first military rule of Niger. While the Supreme Military Council did release the political prisoners of President Diori, they instituted numerous other crack downs on democracy and liberty, banning political parties and harshly ending and punishing attempted coups. Finally, after stable but meager economic growth, in 1989 a referendum created the second republic, which was notable for its liberalization of the legal system, and the release of all previous political prisoners. Though, key to the transition of post-independence Niger, the 1991 National Severing Conference opened up the country into a multi-party democracy, and became the third republic. It was followed not long after by another military coup, which seized power given the unacceptable gridlock perceived by the public in this new multi-party system. Therefore, we're left with the case in Niger between 1989-1998, from the first democratic transfer of power in 1993, to the later coup in 1996, the democratic government was woefully underequipped to make substantive change in that three-year period. In this sense we begin to understand the key point that makes election plateau countries similar, the elected government lacked the sufficient time and power to affect real change, and was ejected before democracy could truly grow. In Niger, we found a standard example of the story of election plateau countries, short rules which resulted in no significant change to democracy or development.

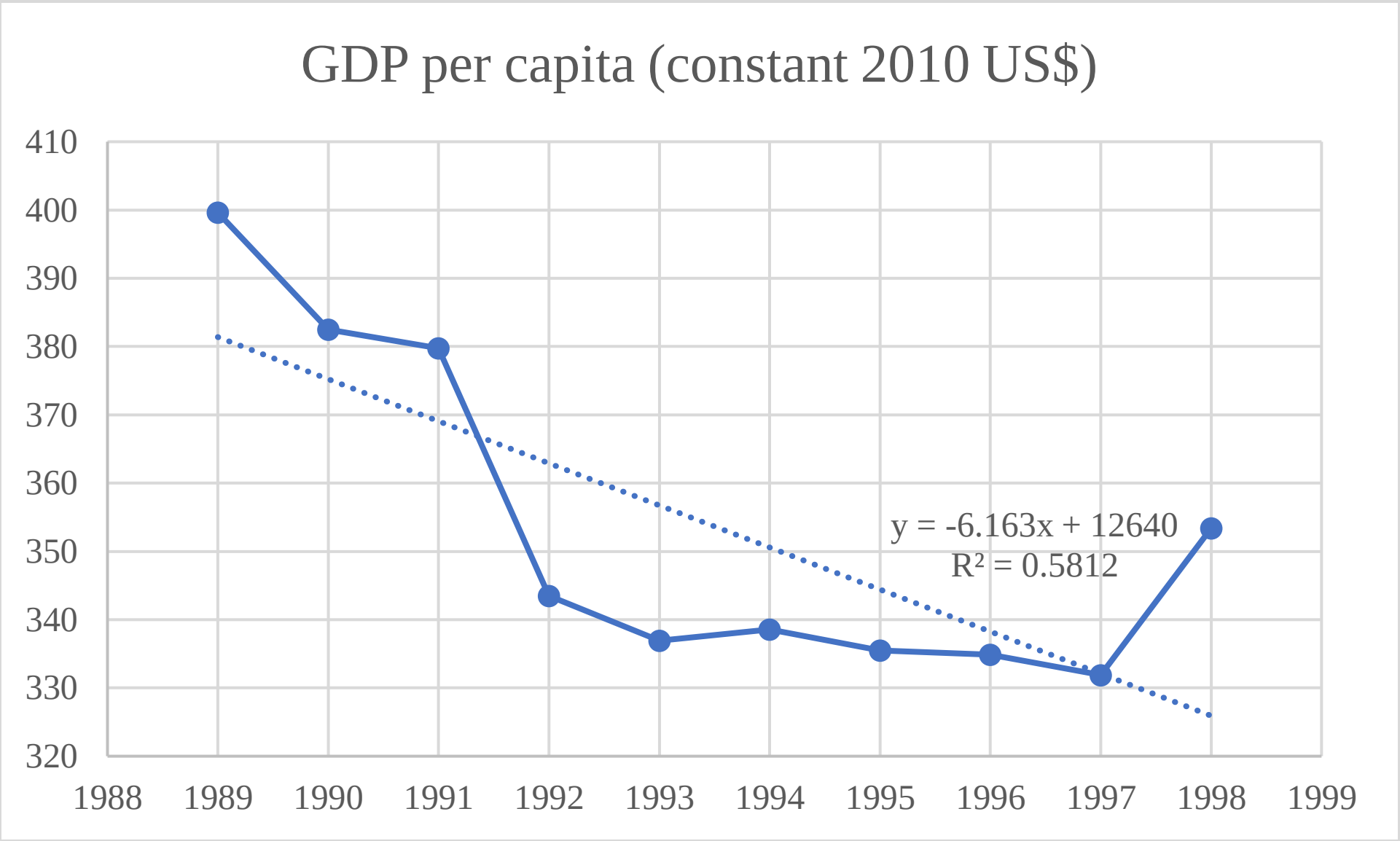

*Figure 1.2*

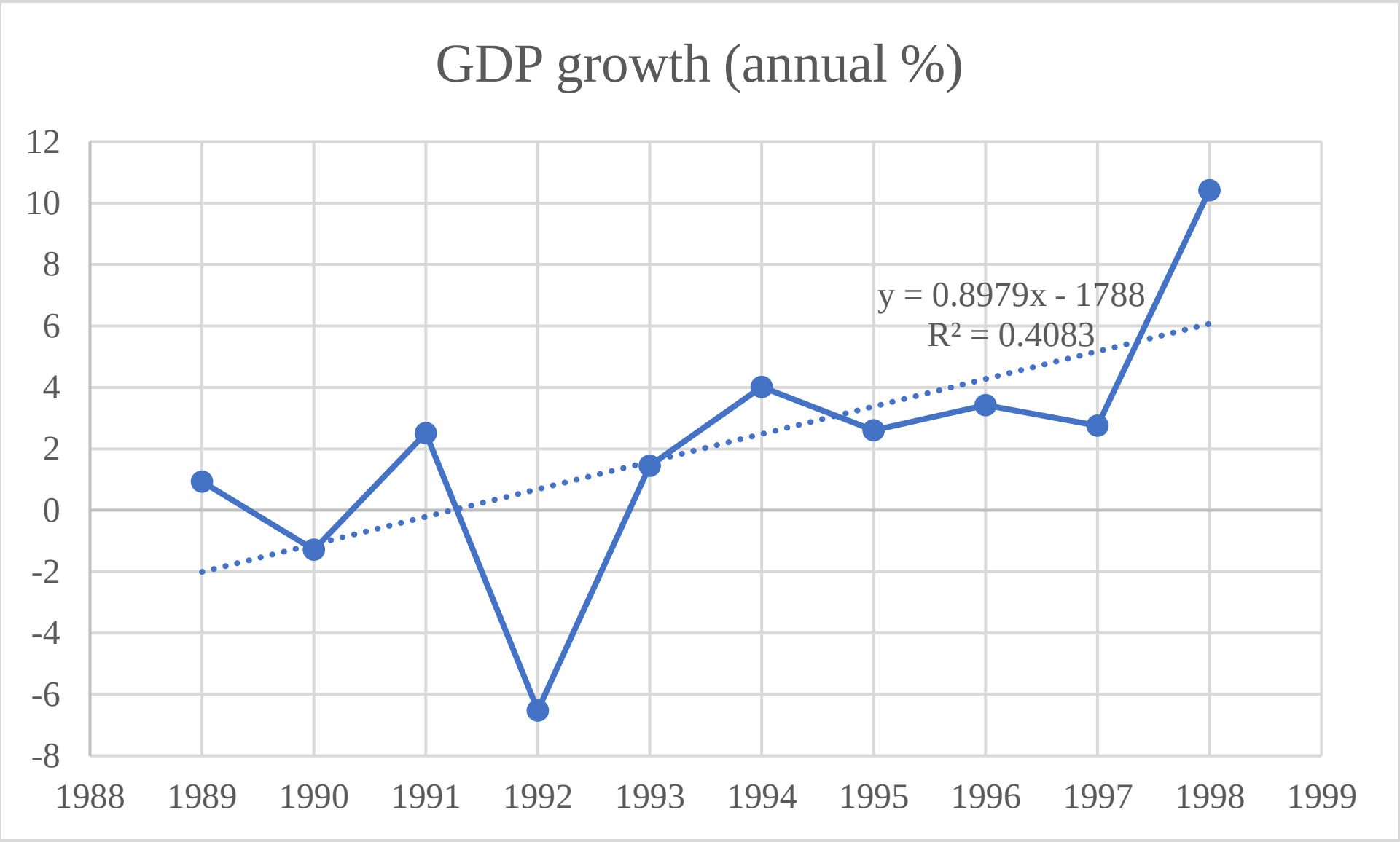

*Figure 1.3*

**Benin**

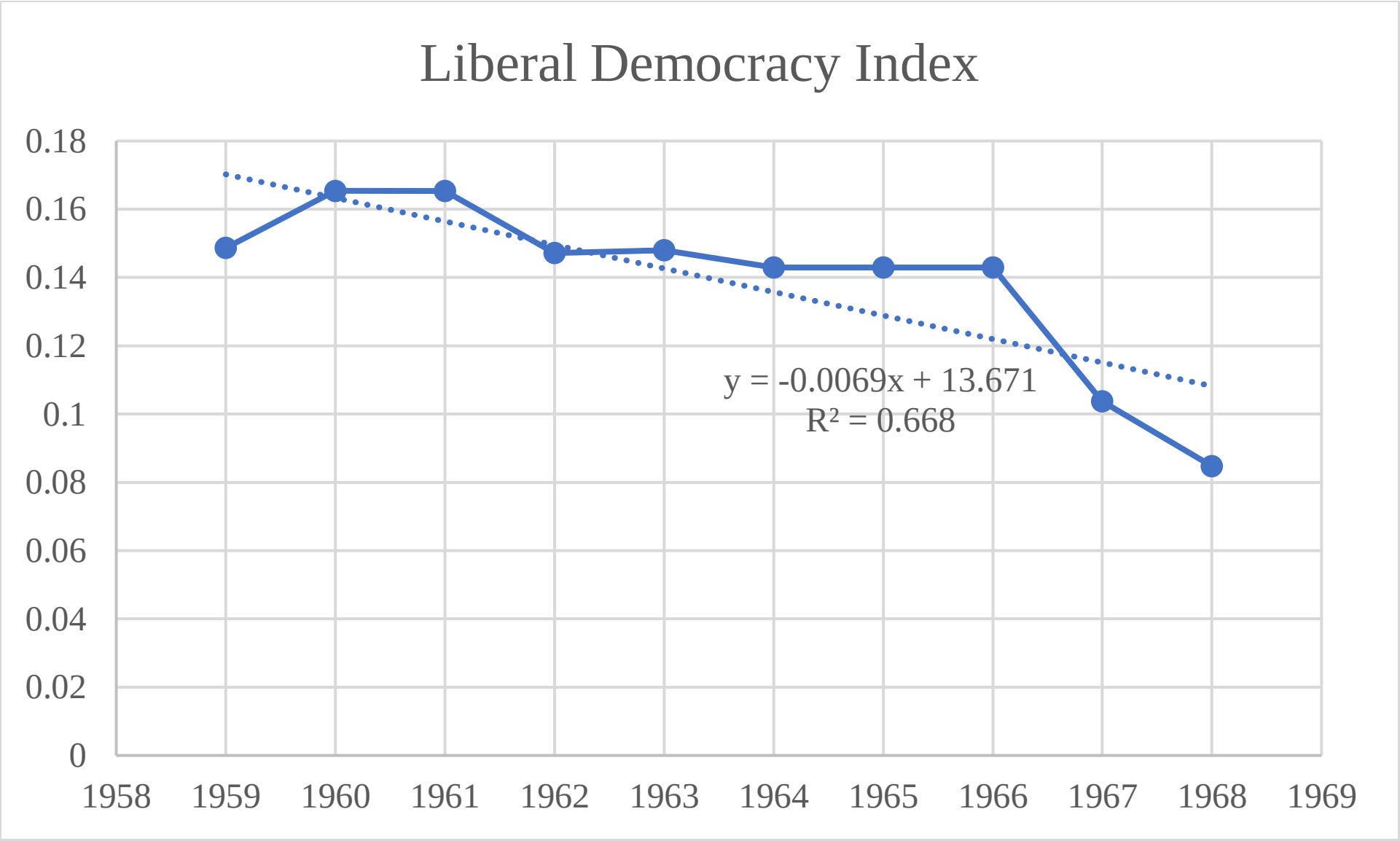

*Figure 2.1*

Benin exemplifies the slope down archetype, various failed coups and multiple seizures of power have plagued the administration with a crippling inability to make meaningful change or to invest properly in the development of the country. This is the story of slope down countries, failures and falling further and further into authoritarianism. This again lends credence to the idea that authoritarianism has an inverse causal link to development. Before 1960, Benin was under French colonial rule. While there was certainly investment in infrastructure, French colonialism in Africa focused on assimilation through education and culture, not through actual marriages between ethnic French and local populations (Betts 2006). This meant that while some infrastructure was developed, since few ethnic French lived there, less 'foreign' wealth and development was brought in. Then, in 1960, The country received its independence, and spent the next decade between various failed elections and military coups. Obviously, in this chaotic, nearly anarchic state, very little investment or development existed to support the country. While, overall GDP may have grown, the negative trend in GDP growth reveals how the lack of oversight, protections for private property, and general safety reveals how the lack of effective democracy has adverse effects on overall development. In Benin, as the country's leadership falters so too does the countries development indices. As the complications around the elections in the years immediately following independence unfolded, we saw serious adverse effects in the liberal democracy index and development scores of the country. Because of our restricted timeframe, We should likely be able to attribute the GDP per capita to independence, but the negative trend in the much more fickle and temporal GDP growth metric were likely caused by the transitions away from democracy.

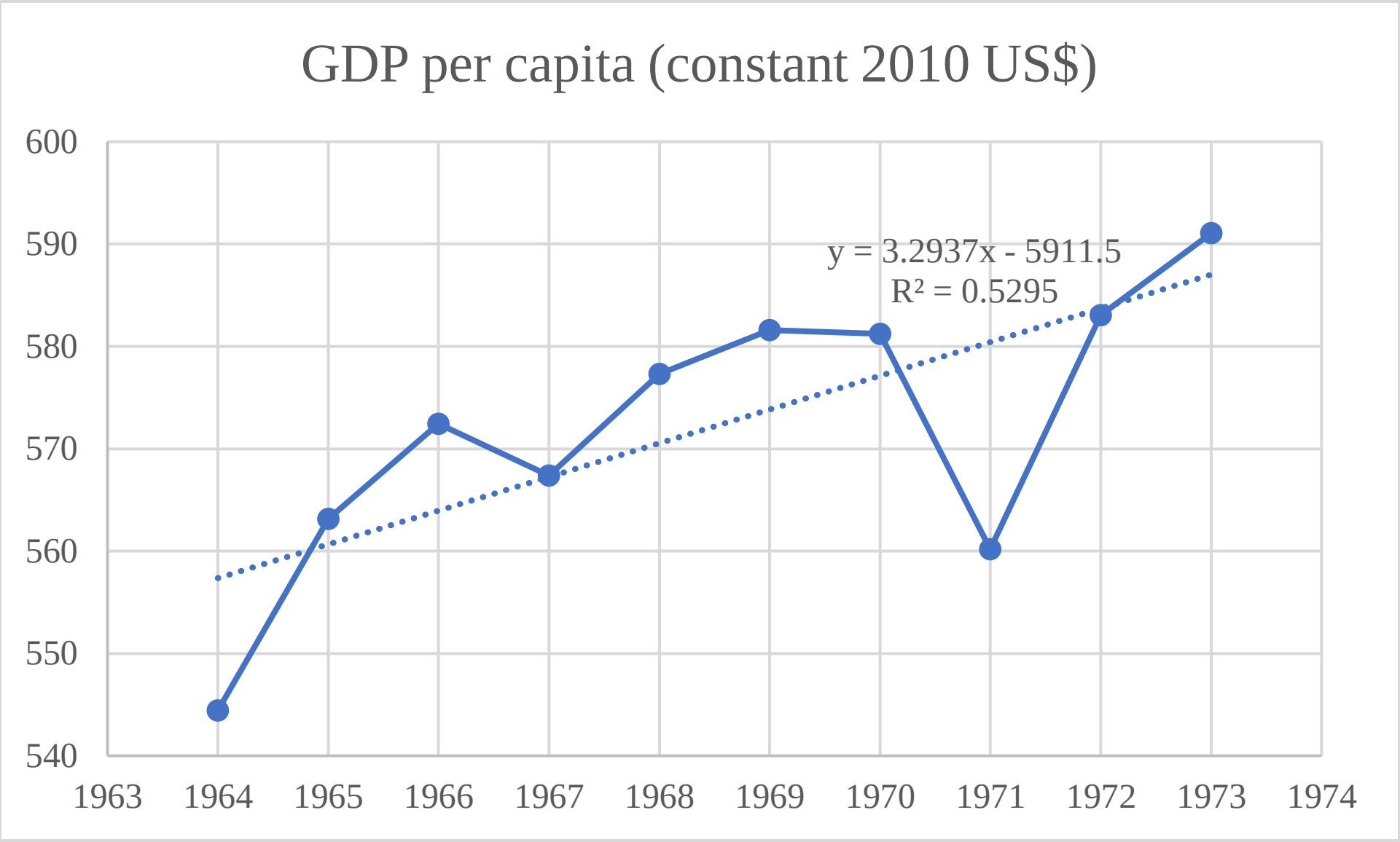

*Figure 2.2*

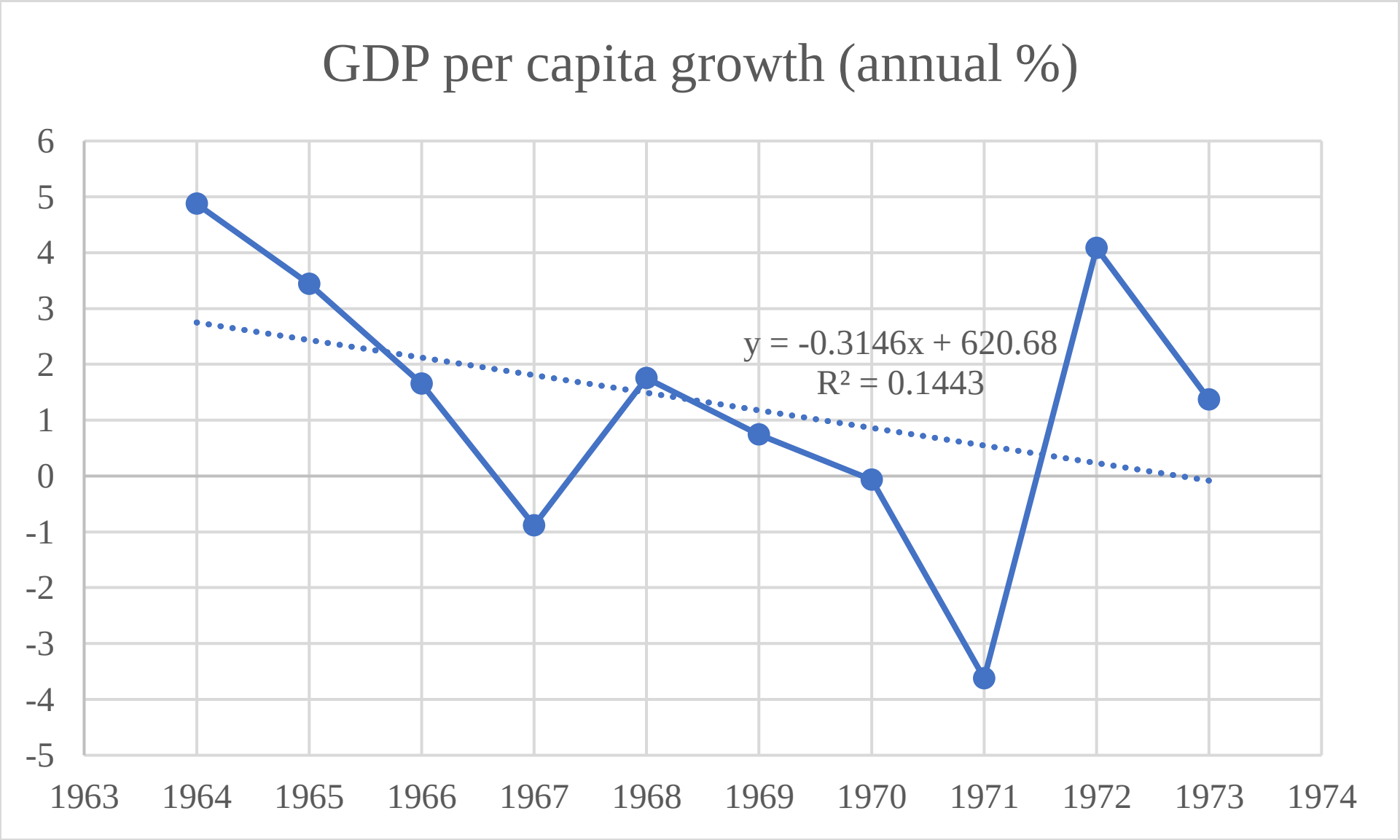

*Figure 2.3*

**Zimbabwe**

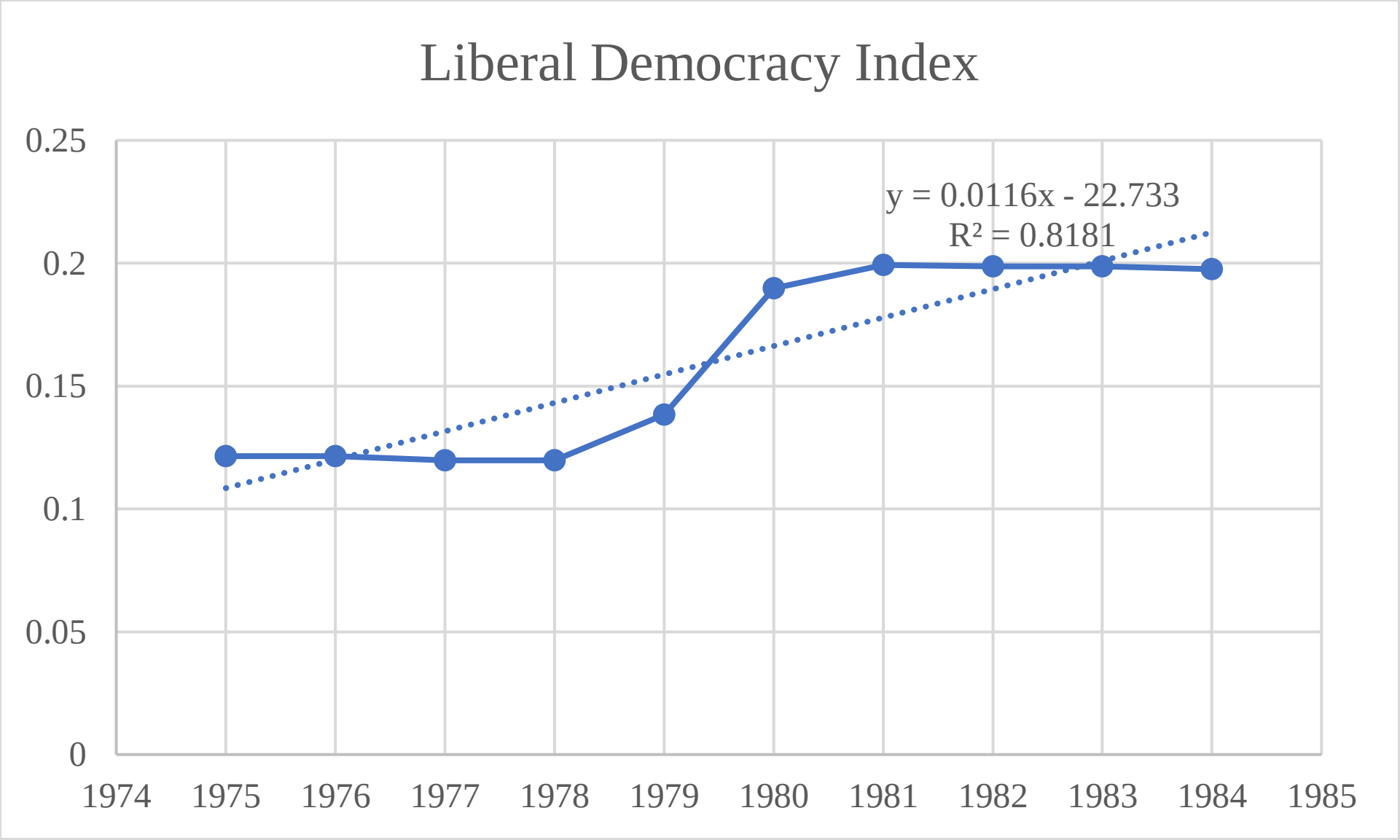

*Figure 3.1*

Zimbabwe is truly the poster child for the study. While the transfer of power in the case studied was from colonial to independent power, it still represented a significant increase in their liberal democracy score, and saw marked improvements in GDP per capita and GDP growth as a result.

As the country celebrated its new-found independence, the surge towards democracy drove development, and brought about improvements in numerous sectors of the economy and public interest. During the celebrations, the then president of Nigeria, Shehu Shagari, pledged $15 million for the training of Zimbabweans, with the majority of those funds supporting students in university (Kalley, Schoeman and Andor 1999). This kind of strong foreign investment, coupled with strong protections for private property fueled the lighting growth of post-independence Zimbabwe. Overall, the first few years of Zimbabwe's independence were characterized by general unity, cohesion, and it tend to enjoy the benefits of a strong liberal democracy, mostly through its unity against the

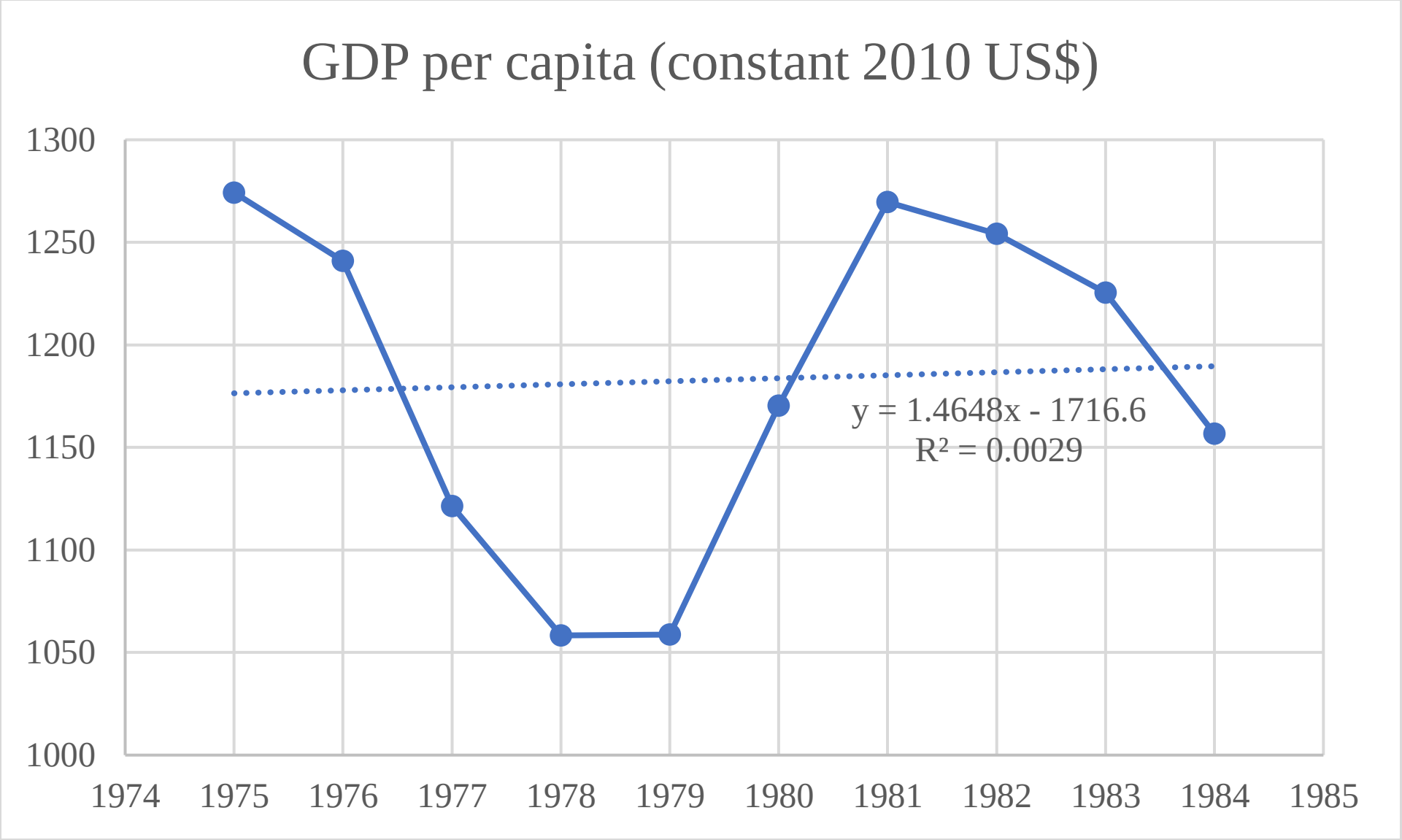

*Figure 3.2*

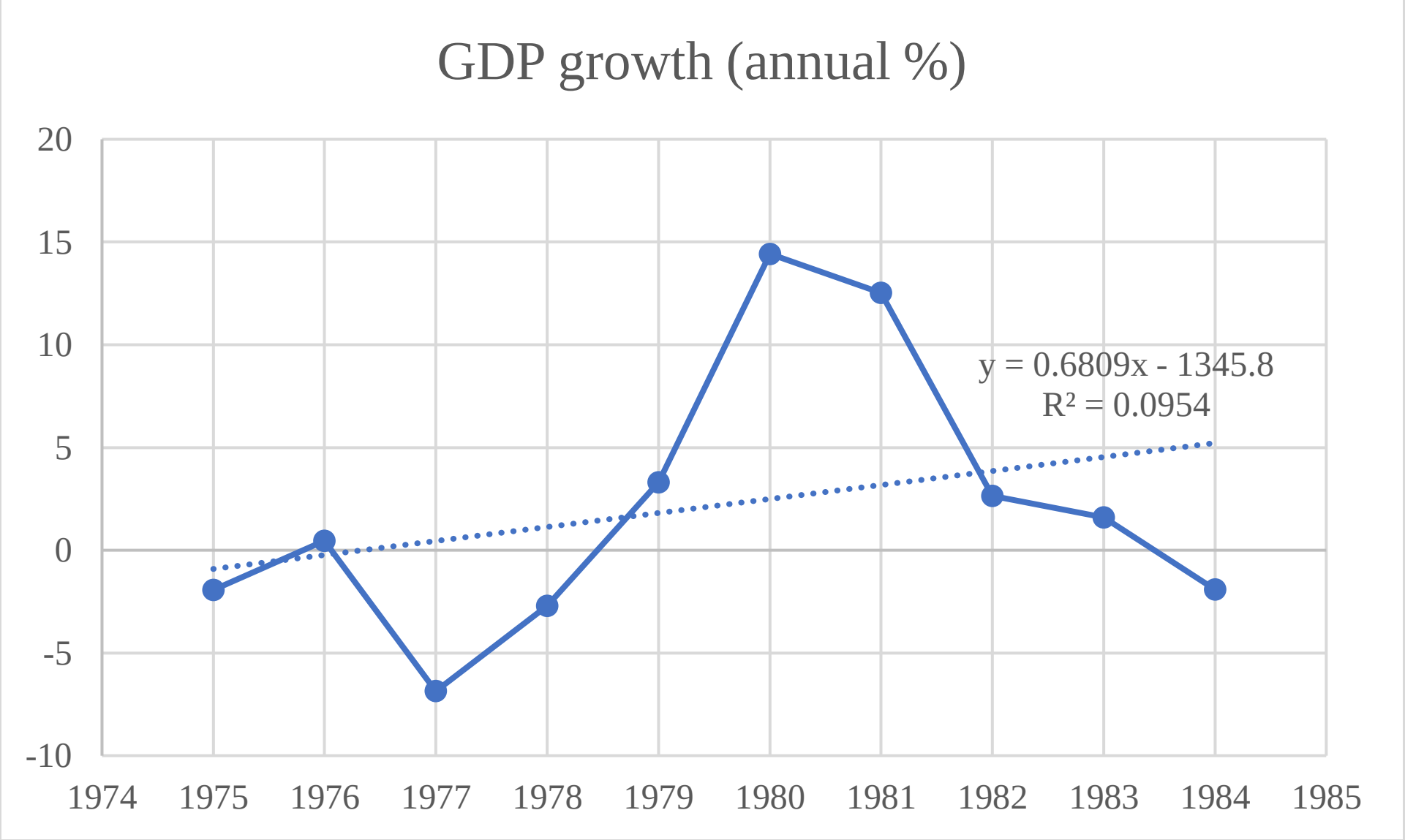

*Figure 3.3*

remnants of British rule rather than real institutional protections of freedoms. While Zimbabwe's independence story is not necessarily the norm, the trend it represents for step up democratic transitions seems clear, improvements in the administration and adherence to democracy seem to more often than not show similar improvements in developments. Later, partisanship, ethnic divisions, and general disunity would fracture the country and mean that more institutional progress was required to better ensure the continued success of the country, but in the short term, immediately following independence and the election of the first independent president, the Zimbabwe saw numerous, concrete benefits to its development.

## Conclusion

In the end, we found numerous archetypes in the liberal democracy indices, but three amongst them had the greatest relevance to our question. First, election plateau helped us learn more about the nuances of transitions to democracy and their effects on development. The results of the studies of election plateau countries showed that even in the short term, there are some benefits to development from democracy, but the fall back into authoritarianism often counteracted much of the benefits gained in the short term, and nullified the possibility of long term benefit. The second, slope down, showed us how authoritarianism often led to negative results in development, for numerous factors. Finally, the last major archetype, step up, showed us the benefits of strengthening liberal democracy on development, though it was difficult to more accurately say which methodologies exactly lead top the benefits democracy displays. Overall, what we found from analyzing these three most indicative archetypes of democratic transitions, was that democracy does appear to lead to increased development. Unfortunately though, the underlying reasons for this have not been clearly shown by this study and numerous questions remain ripe for further analysis.